\documentclass[reprint,aps,prl,superscriptaddress,twocolumn,floatfix]{revtex4-1} 

\usepackage{graphicx}
\usepackage{amsmath}
\usepackage{bm}
\usepackage{color}
\usepackage{soul}
\usepackage{ulem}

\begin{document}

\title{Structure and dynamics of a glass-forming binary complex plasma with non-reciprocal interaction}

\author{Yi-Fei Lin}
\affiliation{College of Science, Donghua University, Shanghai 201620, PR China}
\author{Alexei Ivlev}
\affiliation{Max Plank Institute for Extraterrestrial Physics, Garching 85748, Germany}
\author{Hartmut L\"owen}
\affiliation{Institut f\"ur Theoretische Physik II: Weiche Materie, Heinrich-Heine-Universit\"at D\"usseldorf, 40225 D\"usseldorf, Germany}
\author{Liang Hong}
\affiliation{School of Physics and Astronomy, Shanghai Jiao Tong University, 200240 Shanghai, PR China}
\author{Cheng-Ran Du}
\email{chengran.du@dhu.edu.cn} \affiliation{College of Science, Donghua University, Shanghai 201620, PR China}
\affiliation{Member of Magnetic Confinement Fusion Research Centre, Ministry of Education, PR China}

\begin{abstract}
In this letter, we present the first numerical study on the structural and dynamical properties of a quasi-two-dimensional (q2D) binary complex plasma with Langevin dynamics simulation. The effect of interaction with non-reciprocity on the structure is investigated by comparing systems with pure Yukawa and with point-wake Yukawa interactions. The long-time alpha-relaxation for the latter system is revealed by plotting and analyzing the intermediate scattering function. The results clearly indicate that a q2D binary complex plasma is a suitable model system to study the dynamics of a glass former. The non-reciprocity of the interactions shifts the glass formation significantly but leads to the same qualitative signatures as in the reciprocal case.
\end{abstract}

\maketitle

{\it Introduction.} A complex plasma consists of micron-sized particles immersed in a low temperature plasma \cite{Fortov:2005}. The particles interact with ions and electrons and acquire high charges. Since the thermal velocity of electrons is much higher than ions, the charges $Q$ usually are negative. The particles interact with each other via a screened Coulomb interaction. The interplay between the damping (caused by a neutral gas) and the heating mechanism (such as charge fluctuation) results in a finite kinetic temperature $T$. In a complex plasma, the microparticles are the dominant species in terms of the energy and momentum transport and therefore can be seen as a one-species media. The thermodynamics properties for a one-component complex plasma can be characterized by two dimensionless parameters \cite{Hamaguchi:1997,Hartmann:2005}:
\begin{equation}
  \kappa=\frac{a}{\lambda} \quad \text{and} \quad \Gamma=\frac{Q^2}{4 \pi \epsilon_0 a k T},
\end{equation}
where $a$ is the mean interparticle distance and $\lambda$ is the effective screening length. Complex plasmas represent an open non-Hamiltonian system and can exist in gaseous, liquid and solid forms. Therefore, they are ideal model systems to study classical condensed matter at the ``atom'' (i.e. particle resolved) level \cite{Morfill:2009,Chaudhuri:2011}.

Since the discovery of plasma crystals \cite{Thomas:1996,Chu:1994}, complex plasmas have been used to study the dynamics of various phenomena such as melting \cite{Zuzic:2006,Nosenko:2009}, recrystallzation\cite{Knapek:2007}, defect transport \cite{Nosenko:2007}, formation of Mach cones \cite{Samsonov:2000,Du:2012}, etc. Particularly in laboratory experiments, the particles are suspended in the sheath area above the bottom electrode and form a layer of particles. Due to the convenience of diagnostics, many experiments have been performed in this system.

As a liquid is cooled, it may not only crystallize but also remain in an amorphous state, depending on the cooling rate and the complexity of the liquid \cite{Debenedetti:2001,Berthier:2011}. A supercooled liquid exhibits glassy dynamics. The short-time in-cage motion is described by beta-relaxation, while the long time structural relaxation is called alpha-relaxation. Recently, a glass former was discovered experimentally in a quasi-two-dimensional (q2D) binary complex plasma \cite{Du:2016}. As a ``plasma state'' of the soft matter, the dynamical time scales  in a complex plasma are stretched to tens of millisecond. With a dilute background gas, the system can be seen as virtually undamped \cite{Morfill:2009}. This enables us to study the dynamical property of a glass former from the Newtonian behavior of in-cage motion crossover to the fully damped Brownian dynamics at long timescale \cite{Du:2016}.

In this Letter, we report on a Langevin dynamics simulation of a q2D binary complex plasma. We study the dependence of the structure on the number ratio of two particle types and focus on the dynamics of the long time scale, namely the alpha-relaxation. Fitting the self-part of the intermediate scattering function using a Kohlrausch equation, we determine the critical parameters for the glass transition of the system.

The implications of this study are two-fold: first we show that in an appropriate model for a binary complex plasma a glass transition does occur. Hence complex plasmas can be used to explore the slowing down in
the dynamics. Second, on a more fundamental level, the effect of non-reciprocal interactions on the glass transition has not yet been studied. Computer simulation so far assumed reciprocal interactions and
mode coupling theory typically requires Hamiltonian systems with reciprocal interactions. Our study reveals that non-reciprocal interactions shift the glass transition relative to that in reciprocal systems but the qualitative scenario and signatures are the same.

{\it Simulation.} In complex plasmas, the motions of microparticles can be studied by molecular dynamics (MD) simulation. The equation of motion including damping reads
\begin{equation}
m_i\ddot{\bm{r}}_i+m_i{\nu}\dot{\bm{r}}_i=\sum_j\bm{F}_{ji}+\bm{L}_i,
\end{equation}
where $\bm{r_i}$ is the particle position, $m_i$ the mass, $\nu_i$ the damping rate, $\bm{L}_i$ the Langevin heat bath. The Langevin force $\bm{L}_i$ is defined by $<\bm{L}_i(t)>=0$ and $< \bm{L}_i(t)\bm{L}_i(t+\tau')>=2{\nu_i}{m_i}T\delta(\tau')\bf{I}$, where $T$ is the temperature of the heat bath, $\delta(\tau')$ is the delta function and $\bf{I}$ is the unit matrix.

\begin{figure}[!ht]
  \begin{center}
    \includegraphics[width=0.46\textwidth, bb=0 20 520 205]{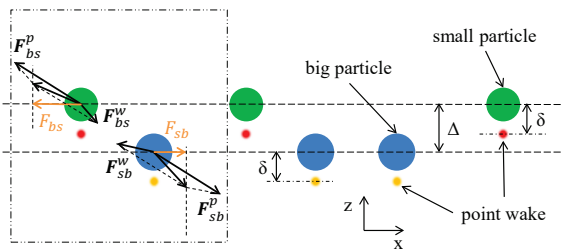}
    \caption{(Color online) A sketch for the side view of a quasi-two-dimensional complex plasma. The non-reciprocal interaction between a small and a big particle is illustrated in the dashed-dotted line rectangle.}
    \label{sketch}
  \end{center}
\end{figure}

Here we simulated a binary mixtures of microparticles confined in the (pre)sheath of a plasma, where the gravitational force is balanced by the electrostatic forces of the sheath field \cite{Hartmann:2009,Du:2016,Wieben:2017}. Since two types of particle have different mass, the equilibrium heights deviate. Generally, this vertical strong confinement can be modeled using parabolic confinement. However, recent experiments show that under certain conditions, the vertical motion is much smaller compared with the horizontal one \cite{Du:2016}. Therefore, for simplicity we neglected the vertical motion and defined the height difference as a constant $\Delta$, see Fig.~\ref{sketch}.

In the simulation, individual microparticles are modeled as negative point-like charges. To include the ion wake in the particle interaction, a positive point-like charge is placed at a fixed vertical distance $\delta$ below each particle \cite{Laut:2017,Yazdi:2015,Zhdanov:2009,Roecker:2012}. The force exerted on particle $i$ by particle $j$ is composed of two components, namely the repulsive force $\bm{F}_{ji}^p$ by particle $j$ and attractive force  $\bm{F}_{ji}^w$ by the point-like charge below particle $j$. Both components have a form of Yukawa interaction, see Fig.~\ref{sketch}. The effective forces can be written as
\begin{equation}
\bm{F}_{ji}=\bm{F}_{ji}^p+\bm{F}_{ji}^w=Q_iQ_jf(r_{ji})\frac{\bm{r}_{ji}}{r_{ji}}+q_iQ_jf(r_{ji}^w)\frac{\bm{ r}_{ji}^w}{r_{ji}^w},
\label{eq_force}
\end{equation}
where $f(r)=\exp(-r/\lambda)(1+r/\lambda)/r^2$, $\lambda$ is the screening length, $\bm{r}_{ji}=\bm{r}_i-\bm{r}_j$ and $\bm{r}_{ji}^w=\bm{r}_i-(\bm{r}_j-\delta\bm{e}_z)$.
Without the vertical motion, we projected forces in $xy$ plane and the simulation becomes essentially 2D. Thus the interaction is non-reciprocal \cite{Melzer:1999,Ivlev:2015}. The  molecular dynamics simulations were performed using LAMMPS in NVT ensemble \cite{lammps,Plimpton:1995}.

\begin{figure}[!ht]
  \begin{center}
    \includegraphics[width=0.46\textwidth, bb=180 330 440 470]{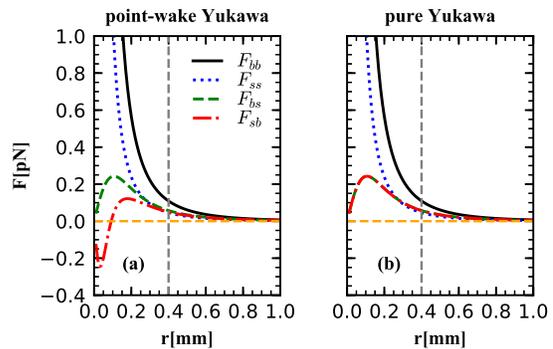}
    \caption{(Color online) The force between two particles with (a) and without (b) considering the wake effect. The positive value means repulsive and the negative value means attractive, the force exerted on particle $i$ by particle $j$ is denoted by $F_{ji}$. The yellow horizontal dashed line marks the zero force and the grey vertical dashed line highlights the screening length. Note that $F_{sb}$ is non-monotonic in pure Yukawa case since particles are in different layers.}
    \label{force}
  \end{center}
\end{figure}

We choose typical plasma and particle parameters according to the experiments \cite{Du:2016}. The neutral gas pressure is $0.065$~Pa (which determines the damping) and the screening length is $\lambda=400$~$\mu$m. The small particles have a diameter, $d_s=9.19$~$\mu$m, and a mass, $m_s=6.13 \times 10^{-13}$~kg. The big particles have a diameter, $d_b=11.36$~$\mu$m, and a mass, $m_b=8.03 \times 10^{-13}$~kg. The charge are set as $Q_s=7280$~$e$ and $Q_b=11200$~$e$ and the damping rates are $\gamma_s=0.77$~s$^{-1}$ and $\gamma_b=0.91$~s$^{-1}$, respectively. The small particles are levitated by $160$~$\mu$m higher than the big particles. For the point-wake model, we fix the vertical distance of the point wake to the particle as $40$~$\mu$m and the point-wake charge is $20$\% of the particle charge \cite{Laut:2017}.

The resulting interaction in the horizontal direction with the point-wake model is shown in Fig.~\ref{force}(a). As we can see in the figure, the interaction between the same type of particles is always repulsive. Due to the higher charge, the repulsive force between big particles $F_{bb}$ is larger than that between small particles $F_{ss}$. However, the force acting on a small particle by a big particle and its wake ($F_{bs}$) is larger than the force on a big particle by a small particle and its wake ($F_{sb}$). Moreover, at very small distance, the later shows an attractive force due to the presence of the wake. In the point-wake model, the interaction is indeed non-reciprocal. For comparison, we show the interaction without considering the wake effect (only $F_{ji}^p$ term in Eq.~\ref{eq_force}) in  Fig.~\ref{force}(b). The interaction is then reciprocal.

We select the kinetic temperature $T$ and $\kappa$ as control parameters. To mitigate the influence of the initial condition, we fix the spatial distribution of particles scaled to the desired interparticle distance. For each pair of control parameters, at least four different initial distributions are probed. It turns out that the initial condition has marginal influence on the relaxation and structural properties.

\begin{figure}[!ht]
  \begin{center}
    \includegraphics[width=0.45\textwidth, bb=190 330 420 470]{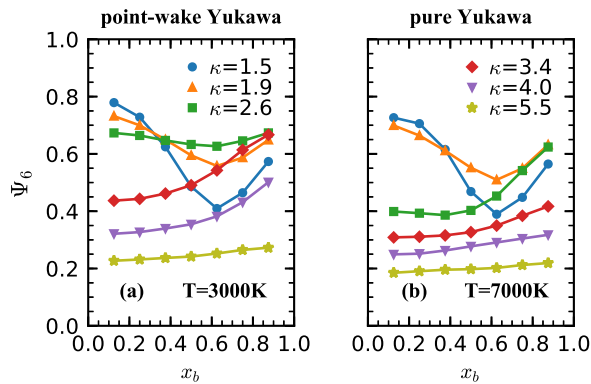}
    \caption{(Color online) Dependence of averaged hexatic order parameter $\Psi_6$ on the number ratio of big particles $x_b$  with (a) and without (b) considering the wake effect. For the point-wake model and Yukawa interaction, the selected temperature is $3000$~K and $7000$~K, respectively. }
    \label{ratio}
  \end{center}
\end{figure}

{\it Results.}
The complexity of system is key for the glass former. For an one-component complex plasma, particles form plasma crystal at certain temperature as $\kappa$ decreases below a critical value. Binary mixture prevents crystallization. The local structure of a complex plasma can be characterized by a time average of local transient hexatic order parameter $\psi_6^i$
\begin{equation}
\bar\psi_6^i=\frac{1}{\tau_a}\int_{t'}^{t'+\tau_a}dt|\psi_6^i| \quad \text{and} \quad \psi_6^i=\frac{1}{n_i}\sum_m e^{j6\theta_m^i},
\end{equation}
where $n_i$ is the number of nearest neighbors of particle $i$ and $\theta_m^i$ is the angle between ${\bf r}_m-{\bf r}_i$ and the $x$~axis projected in the $xy$~plane. We select $\tau_a=10$~s for the average. For a system composed of many particles, we average $\psi_6^i$ over the all the particles included in the system and obtain a structure parameter $\Psi_6=\sum_i\bar\psi_6^i/N$, where $N$ is the number of particles.  Here $\Psi_6=1$ means a perfectly ordered crystal with hexagonal structure, while $\Psi_6=0$ means random arrangement.

In Fig.~\ref{ratio}(a), we plot $\Psi_6$ versus number ratio of two particle species $x_b=n_b / (n_s+n_b)$ at $T=3000$~K. For big $\kappa$ ($>4.0$), $\Psi_6$ increases as the percentage of big particles increases. This slow increase is caused by the increase of the coupling parameters since the big particles have more charges. However, for all these $\kappa$ and $x_b$, $\Psi_6$ is below $0.6$. The structure is quasi-random. As $\kappa$ further decreases to $3.4$, the crystal structure emerges as the coupling increases. For $\kappa=2.6$, the system shows a ordered structure regardless the ratio of mixture. For strongly coupled system, as shown by $\kappa=1.9$, $1.5$, the system exhibits ordered structure as either species dominates regardless of the coupling strength. However, for the two types of particles with similar portions, the crystalline order is suppressed. The minim is close to $x_b \approx 0.6$. For simplicity, we use a binary mixture with $x_b=0.5$ for the study.

For comparison, we test the dependence of the order parameter on the ratio without considering the wake effect. At $T=3000$~K, the system has ordered structure regardless of the ratio for all the selected $\kappa$. Similar trend emerges if we increase the kinetic temperature to $T=7000$~K, as we show in Fig.~\ref{ratio}(b). For small $\kappa$, the crystal structure is suppressed as comparable portion of two types of particles are mixed.

\begin{figure}[!ht]
  \begin{center}
    \includegraphics[width=0.47\textwidth, bb=190 290 410 500]{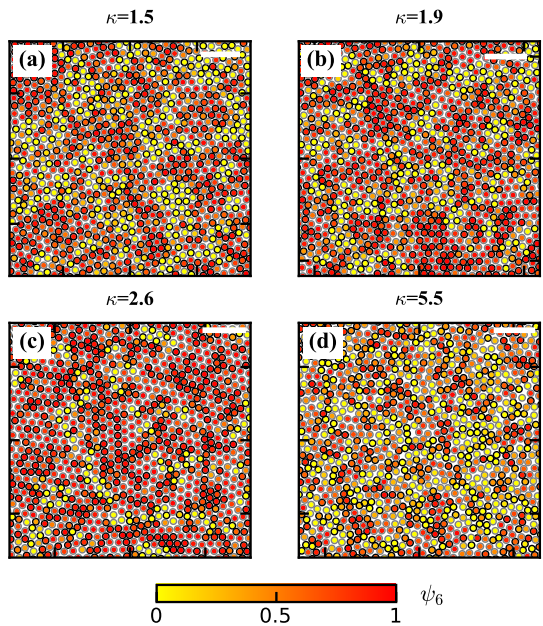}
    \caption{(Color online) Snapshots of particle positions in the q2D binary complex plasma with point-wake Yukawa interaction. Big particles have a black outlines while the small particles have a grey outlines. The filled color coded from yellow to red shows the $\Psi_6$ from $0$ to $1$. The temperature is $1000$~K. The white bar shows ten inter-particle distances.}
    \label{snapshot}
  \end{center}
\end{figure}

We show the spatial distribution of $\psi_6$ at different $\kappa$ in Fig.~\ref{snapshot}. As one can see, at $T=3000$~K small and big particles are homogeneously distributed in the area regardless of coupling strength. For small $\kappa$, the local structure of most particles is random though one can see clusters of hexagonal structure (yellow blobs). The size of the such cluster increases as $\kappa$ increases until the majority particle are arranged in a hexagonal structure. However, as the coupling further decreases, particles start to arrange randomly again, as shown in Fig.~\ref{snapshot}(d). This trend agrees with Fig.~\ref{ratio}(a) for $x_b=0.5$.

\begin{figure}[!ht]
    \includegraphics[width=0.47\textwidth, bb=200 340 420 470]{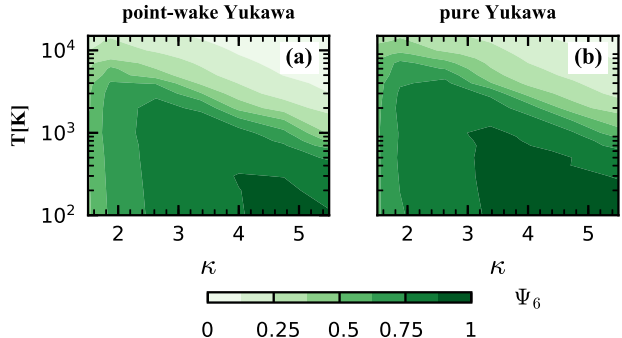}
    \caption{(Color online) The averaged hexatic order parameter $\Psi_6$ in the ($T$,$\kappa$) plane, for a binary complex plasma with (a) and without (b) considering the wake effect.}
    \label{structure}
\end{figure}

We plot a diagram of averaged hexatic order parameter versus $T$ and $\kappa$ to gain a general picture, as shown in Fig.~\ref{structure}(a). As the system cools down, the fast motion of particles decreases and the order starts to build up. As we can see in the figure, the lower the temperature is, the more ordered the particles are. For small $\kappa$, the structure doesn't change much for low temperature but hexagonal order builds up as temperature decreases in the ``hot'' regime. For big $\kappa$, temperature of this fast changing regime is substantially lower. Note that for a certain temperature, the change of the structure is not monotonic against $\kappa$. There is only a window of $\kappa$ in which the system forms a hexagonal ordered structure.

Again we plot the diagram of hexatic order parameter for a Yukawa system without considering the wake. As we can see in Fig.~\ref{structure}(b), for the same pair of $T$ and $\kappa$, $\Psi_6$ has a higher value, representing ordered structure. This is caused by the higher coupling if the wake is absent. However, the hexagonal structure is formed in a larger parameter window in the $T$-$\kappa$ diagram. This shows that for the reciprocal and non-reciprocal system, the formation of ordered structure has a qualitatively similar trend in the parameter range of interest in a binary complex plasma.

The structural relaxation is generally quantified by the density-density correlation function in ${\bf q}$-space, $F(q,t)$, which is the Fourier-transformation of the van Hove correlation function \cite{Hansen:book}, commonly referred to as the intermediate scattering function (ISF). For practical purposes, it is convenient to use the self-part of ISF to describe the evolution of single-particle correlations
\begin{equation}
F_{\rm s}(q,t) = N^{-1}\langle\sum_i^N\exp \{-i{\bf q}\cdot[{\bf r}_i(t+t_0) - {\bf r}_i(t_0)]\}\rangle,
\end{equation}
where ${\bf r}_i(t)$ is the position of the particle $i$ at the moment $t$, and $\langle\ldots\rangle$ denotes the canonical average (over $t_0$). ${\bf q}$ is the wave number and we selected $|{\bf q}|=\pi/\Delta$  \cite{Fuchs:1998,Voigtmann:2004,Bayer:2007,Flenner:2015,Kawasaki:2007,Vivek:2017}.

The stretched-exponential (Kohlrausch) law \cite{Franosch:1997,Fuchs:1992,Fuchs:1994,Hansen:book,Voigtmann:2004,Feng:2010b},
\begin{equation}
F_s(q,t)\simeq A(q)\exp\{-[ t/\tau(q) ]^{\beta(q)}\}
\end{equation}
usually provides a good fit for the long-time asymptote of ISF, the so-called alpha-relaxation. The law is determined by three parameters: the amplitude factor $A(q)$, the timescale of the alpha-relaxation $\tau(q)$, and the stretching exponent $\beta(q)<1$. Selecting a time domain appropriate for the fit is generally not an easy task \cite{Fuchs:1992,Bartsch:1992,Phillips:1996,Voigtmann:2004} -- an overlap with the transient beta-relaxation should be avoided, which imposes the lower time limit for the fit.

\begin{figure*}[!ht]
  \begin{center}
    \includegraphics[width=0.72\textwidth, bb=60 270 540 530]{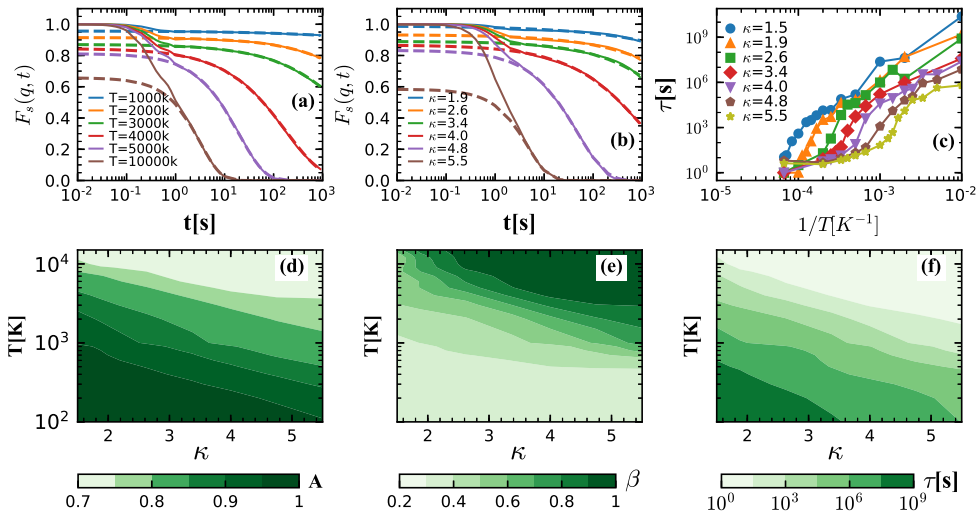}
    \caption{(Color online) Intermediate scattering function (ISF) for varying temperature at $\kappa=2.6$ (a) and for varying $\kappa$ at $T=2000$~K (b) in a q2D binary complex plasma. The wave number $q$ is set equal to $\pi/\Delta$. For selected $\kappa$, the relation between the fitted $\tau$ and $1/T$ is plotted in (c). The ISFs are fitted by Kohlrausch law, the lower panel shows the fitting parameter $A$ (d), $\beta$ (e), and $\tau$ (f) in the ($T$,$\kappa$) plane.}
    \label{dynamics}
  \end{center}
\end{figure*}

In Fig.~\ref{dynamics}(a) we plot the ISF of a system with wake-point Yukawa interaction for different temperature ranging from $1000$ to $10000$~K at $\kappa=2.6$. For $T=1000$~K, $F_s$ shows a small drop at $t \approx 0.8$~s, representing the beta-relaxation. It then reaches a plateau without significant drop within the time of numerical experiment. As temperature increases, one can see clearly the long-time alpha-relaxation, separated by the plateau from the beta-relaxation. The system goes through the certain glass transition. For very high temperature ($T=10000$~K), the plateau disappear and two steps of relaxation starts to merge. To complement Fig.~\ref{dynamics}(a), we fix a kinetic temperature ($T=2000$~K) and plot the ISFs for various $\kappa$ in Fig.~\ref{dynamics}(b). Similarly, we see the plateau between alpha- and beta-relaxation for small $\kappa$, which vanishes for big $\kappa$.

It is instructive to look into the relation between $\ln(\tau)$ and $1/T$. As we see in  Fig.~\ref{dynamics}(c), for all $\kappa$, $\ln(\tau)$ first increases slowly with $1/T$ and the increase becomes steep. For small temperature, this increase becomes slow again.

The Kohlrausch amplitude $A$, the characteristic relaxation time $\tau$, and the stretching exponents $\beta$ are plotted in $T$-$\kappa$ space in Fig.~\ref{dynamics}(d,e,f), respectively. As we can see, the relaxation time decreases dramatically with interpaticle distance and kinetic temperature. $A$ decreases from $1$ to $0.7$ as $\kappa$ increases from $2$ to $5$ and temperature increases from $100$ to $10000$~K. The stretching exponents show a similar trend.

{\it Conclusion} Using computer simulations we have explored the glass transition in a binary model appropriate for a binary dusty plasma sheet with non-reciprocal interactions induced by the presence of a wake charge. As a reference, we have computed also the corresponding behavior of a Yukawa system without wake charge with reciprocal interactions. Non-reciprocal interactions shift the location where the dynamics is slowing down significantly but do not change the qualitative signature of the glass transition. This is interesting as a priori one could have expected that the local heating induced by non-reciprocity have have lead to a different dynamical cross-over. For the future it would be interesting to establish a connection between our kind of non-reciprocity and that occurring in active systems where jamming and vitrification has also been found recently \cite{Flenner:2016,Berthier:2017,Nandi:2017,Janssen:2017,Liluashvili:2017}.

\begin{acknowledgments}
The authors acknowledge support from the National Natural Science Foundation of China (NSFC), Grant No. 11405030. We are thankful for support of this work by the Deutsche Forschungsgemeinschaft (DFG) through the grants IV 20/3-1 (for A.I.) and LO 418/23-1 (for H.L.).
\end{acknowledgments}


%

\end{document}